\documentclass[conference]{IEEEtran}
\IEEEoverridecommandlockouts

\usepackage{cite}
\usepackage{amsmath,amssymb,amsfonts}
\usepackage{algorithm}
\usepackage{algorithmic}
\usepackage{graphicx}
\usepackage[caption=false,font=footnotesize]{subfig}
\usepackage{textcomp}
\usepackage{xcolor}
\usepackage{braket}
\usepackage{booktabs}
\usepackage{array}
\usepackage{url}
\usepackage[hidelinks]{hyperref}
\usepackage{tikz}
\usetikzlibrary{arrows.meta,positioning,fit}
\usepackage{geometry}
\def\BibTeX{{\rm B\kern-.05em{\sc i\kern-.025em b}\kern-.08em
    T\kern-.1667em\lower.7ex\hbox{E}\kern-.125emX}}

\begin{document}

\title{Vehicle-Routing Clustering by Sparse Quantum Relaxation}

 \author{\IEEEauthorblockN{Farzan Moosavi}
 \IEEEauthorblockA{\textit{Laboratory of Innovations in Transportation (LiTrans)} \\
 \textit{Toronto Metropolitan University}\\
 Toronto, Canada \\
 0009-0008-8490-3851}
 \and
 \IEEEauthorblockN{Bilal Farooq}
 \IEEEauthorblockA{\textit{Laboratory of Innovations in Transportation (LiTrans)} \\
 \textit{Toronto Metropolitan University}\\
 Toronto, Canada \\
 0000-0003-1980-5645}}
\maketitle

\begin{abstract}
This paper studies the cluster-assignment layer of a cluster-first-route-second decomposition for the capacitated pickup-and-delivery problem with time windows (CPDPTW), instead of a direct route-ordering quadratic unconstrained binary optimization (QUBO). Sample-based subspace diagonalization (SQD) degenerates to best-of-shots for any diagonal optimization Hamiltonian, so a useful transfer to routing requires non-diagonal relaxation. Quantum random access optimization (QRAO) provides this non-diagonality by encoding binary variables through non-commuting quantum random access code (QRAC) observables. CPDPTW-derived conflict graphs expose a routing-specific obstruction because dense objectives and penalty-encoded constraints destroy QRAO compression. We keep the Hamiltonian sparse by encoding only the sparsified objective and enforcing one-hot, capacity, and cluster time-budget feasibility by repair after rounding. We make no quantum-advantage claim. Under strong repair the classical pass alone reaches near-optimal cost at the tested sizes. The result is a regime map linking sparsification, achieved compression, linear-combination-of-unitaries (LCU) 1-norm, heavy-hex two-qubit depth, sampling drift, and final routing quality on Qiskit Aer, calibrated \texttt{FakeTorino}, and IBM Heron \texttt{ibm\_quebec}. Across a 108-row ideal-calibrated-hardware evaluation grid spanning four to twelve pickup--delivery request pairs, the hardware slice has mean gap 0.0016 and mean device-vs-ideal total variation distance (TVD) 0.279; a decoder ablation at sizes six through twelve isolates the non-diagonal signal.
\end{abstract}

\begin{IEEEkeywords}
Vehicle routing, quantum sampling, quantum relaxation, QRAO, Max-$K$-Cut, CPDPTW, noisy quantum devices
\end{IEEEkeywords}

\section{Introduction}
\label{sec:intro}

The vehicle routing problem (VRP), introduced for truck dispatching and now central to urban logistics, is a classical NP-hard optimization family~\cite{dantzig1959truck}. This paper studies the capacitated pickup-and-delivery problem with time windows (CPDPTW), which adds request pairing, precedence, capacity, and temporal feasibility to the base VRP. Because route quality drives fleet-kilometers, energy use, and greenhouse-gas emissions across ground and drone last-mile fleets, better CPDPTW solvers help decarbonize urban freight, which motivates exploring any solver family---including quantum---that improves solution quality or wall-clock time at fixed cost. Classical metaheuristics, including adaptive large neighborhood search (ALNS), and learning-based solvers remain strong at practical scales~\cite{ropke2006adaptive,liu2023heuristics,bogyrbayeva2022learning}, but dense real-time request patterns and hardware-limited quantum experiments continue to motivate decomposition-based quantum studies. Quantum computing has therefore been explored for transportation and routing applications~\cite{zhuang2024quantum,alcazar2024enhancing}, often through quantum approximate optimization algorithm (QAOA)-style diagonal cost Hamiltonians~\cite{farhi2014quantum}, quantum-assisted decomposition~\cite{palackal2023quantum,dash2025hierarchical}, or hybrid learning and repair policies~\cite{moosavifarooq2025,moosavi2026rl}.

We use a cluster-first-route-second decomposition in which pickup--delivery pairs are kept together, a maximum $K$-cut (Max-$K$-Cut) or knapsack-like clustering layer creates smaller routing subproblems, and exact or heuristic sequencing solves each cluster. Thus the quantum component studied here is the cluster-assignment surrogate, while the final objective remains the true CPDPTW route cost after classical sequencing. This decomposition is useful because direct route encodings scale poorly in qubit count and circuit depth. Sample-based subspace diagonalization (SQD), including sample-based Krylov quantum diagonalization (SKQD), is a natural candidate to import from quantum chemistry into diagonal QUBO optimization, but for routing that transfer fails unless the Hamiltonian is first made non-diagonal.

The obstacle is the following redundancy result. Let $H_C\ket{x}=C(x)\ket{x}$ be the diagonal Hamiltonian induced by any quadratic unconstrained binary optimization (QUBO), Ising, or Max-$K$-Cut encoding, and let $S$ be the set of sampled bitstrings. The projection of $H_C$ onto $\mathrm{span}(S)$ is diagonal, so its ground state is
\begin{equation}
\arg\min_{\ket{x}\in S} C(x).
\label{eq:best_shots}
\end{equation}
Thus diagonal subspace diagonalization is exactly best-of-shots. For any superposition $\ket{\psi}=\sum_{x\in S}\alpha_x\ket{x}$,
\begin{equation}
\langle \psi|H_C|\psi\rangle=\sum_{x\in S}|\alpha_x|^2C(x)\geq \min_{x\in S}C(x),
\label{eq:convex}
\end{equation}
so no coefficient optimization in the sampled subspace improves on the best sampled configuration. This same convexity explains why expectation-value minimization can average away good samples and why conditional value-at-risk (CVaR) objectives are often better suited to classical optimization Hamiltonians~\cite{barkoutsos2020cvar}.

Non-diagonal relaxation is the remedy. The sampling-based quantum optimization algorithm with quantum relaxation (SQOA-QR) follows this route, using QRAO to build a non-diagonal relaxed Hamiltonian~\cite{sqoa2025}. QRAO maps several classical variables onto one qubit through non-commuting Pauli observables, making subspace diagonalization meaningful~\cite{fuller2024qrao,teramoto2023qrao}. This paper develops that idea for CPDPTW decomposition and identifies the routing-specific obstruction that does not appear in sparse MaxCut benchmarks. Application-derived conflict graphs are dense, and penalty-encoded constraints densify them further. Density directly blocks QRAO compression.

We make five contributions.
\begin{itemize}
\item We state the best-of-shots lemma for diagonal SQD and pair it with a Krylov-dimension sweep on the relaxed Hamiltonian, separating diagonal redundancy from the non-diagonal QRAO signal.
\item Building on the QRAO independent-set requirement~\cite{fuller2024qrao,teramoto2023qrao}, we quantify the density obstruction on routing graphs. Binary-label compression falls from threefold on sparse graphs to no compression on dense CPDPTW instances, and penalty encodings pin interaction density at roughly 0.37.
\item We make repair-after-rounding part of the encoding architecture. Full repair drives all methods to near-zero gap, so decoder-level gains are isolated in the feasibility-only ablation, where SQD-weight reranking helps by an amount that scales with the rerank budget.
\item We give a reproducible characterization protocol with a fixed SKQD-style sampling state, explicit Krylov dimension, ideal Aer, calibrated \texttt{FakeTorino}, and real IBM Heron \texttt{ibm\_quebec} layers.
\item We report calibration and resource metrics on a 108-row ideal-calibrated-hardware grid, including a TVD null floor, an ALNS-prior Spearman check~\cite{moosavifarooq2025}, achieved QRAO compression, LCU 1-norm, heavy-hex depth, and final routing gap.
\end{itemize}

\section{Related Work}
\label{sec:related}

Quantum VRP studies often encode routing directly, leading to $\mathcal{O}(N^2)$-variable formulations, although colored-permutation encodings reduce that overhead to $\mathcal{O}(N\log N)$ qubits~\cite{onah2025colored}. Recent work has therefore considered clustering, warm starts, and quantum-assisted decomposition~\cite{palackal2023quantum,dash2025hierarchical,fraunhofer2024quast}. Related hybrid formulations compare QAOA, quantum singular value transformation (QSVT), and linear-combination-of-unitaries (LCU) circuits inside routing loops, while entropy and noise features can gate shallow quantum repair~\cite{moosavifarooq2025,moosavi2026rl}. Complementary analyses identify quantum-utility requirements for the capacitated VRP (CVRP) under realistic noisy intermediate-scale quantum (NISQ) hardware~\cite{onah2025requirements} and constraint-induced limitations of QAOA in feasible subspaces~\cite{onah2025qaoalimits}. Our question, given that setting, is which Hamiltonian and sampling architecture make the clustering-layer quantum sampler non-redundant.

Sample-based diagonalization is powerful in quantum chemistry because the electronic Hamiltonian is non-diagonal in the determinant basis and the ground state is a genuine superposition~\cite{yu2025skqd,robledo2025sqd,filtersparsity2026}; a diagonal optimization Hamiltonian has every bitstring as an eigenstate. SQOA-QR resolves this mismatch by replacing the diagonal Ising Hamiltonian with a quantum-relaxed one and using sampled subspace methods non-variationally on unconstrained sparse MaxCut~\cite{sqoa2025}, but it does not address constrained routing.

QRAO constructs quantum relaxations using quantum random access codes (QRACs)~\cite{fuller2024qrao}. For a $(3,1)$ QRAC, three classical binary variables can be associated with the non-commuting observables $X$, $Y$, and $Z$ on one qubit. A binary variable $x_i$ is relaxed as
\begin{equation}
x_i \mapsto \frac{1-\sqrt{3}P_i}{2}, \qquad P_i\in\{X,Y,Z\}.
\label{eq:qrao_map}
\end{equation}
The resulting Hamiltonian is non-diagonal whenever non-commuting Paulis appear. Teramoto et al.\ extend the construction with alternative QRACs and show explicit compression--approximation trade-offs~\cite{teramoto2023qrao}. The $(3,1)$-QRAC yields $3\times$ nominal compression with Pauli-rounding worst-case ratio $5/9\approx 0.555$; the $(2,1)$-QRAC yields $2\times$ compression at ratio $5/8=0.625$; the $(3,2)$-QRAC yields $1.5\times$ compression at ratio $0.722$. Nominal compression assumes independent-set packing; achieved compression $\rho_{\mathrm{comp}}(G_s)$ collapses on dense graphs (Section~\ref{sec:density}), so we treat these MaxCut bounds as guidance, not formal guarantees for CPDPTW.

\section{CPDPTW Clustering Surrogate}
\label{sec:formulation}

Consider a CPDPTW instance with pickup--delivery requests $r\in R$, pickup node $p_r$, delivery node $d_r$, positive request size $q_r^+$, vehicle capacity $Q$, and route-time budget $\Theta$. We do not encode the full route-ordering problem on the quantum processor. Instead, each pair $(p_r,d_r)$ is contracted into a supernode $i\in V'$ so that must-link feasibility is enforced before quantum encoding. This avoids a large negative must-link penalty edge, which would inflate coefficient magnitudes and the LCU 1-norm $\lambda$.

The clustering layer assigns each supernode $i\in V'$ to one of $K$ vehicles. With one-hot variables $x_{i,c}\in\{0,1\}$, the incompatibility-weighted Max-$K$-Cut objective is
\begin{equation}
\max_x\sum_{(i,j)\in E}w_{ij}
\left(1-\sum_{c=1}^{K}x_{i,c}x_{j,c}\right),
\label{eq:maxkcut}
\end{equation}
where large $w_{ij}$ means that $i$ and $j$ should be separated. The constraints are
\begin{align}
&\sum_{c=1}^{K}x_{i,c}=1 && \forall i\in V',
\label{eq:onehot}\\
&\sum_{i\in V'}q_i^+x_{i,c}\leq Q && \forall c\in[K],
\label{eq:cap}\\
&\sum_{i\in V'}\delta t_i x_{i,c}\leq \Theta && \forall c\in[K].
\label{eq:tw}
\end{align}
Equations~\eqref{eq:cap}--\eqref{eq:tw} are per-cluster knapsack budgets, not one aggregate inequality. In the experiments $\delta t_i$ is the pickup-to-delivery direct travel time of the contracted pair, so~\eqref{eq:tw} is a cluster time-budget proxy; alternatives (service duration, time-window width, insertion-time proxy) leave the constraint structure unchanged. We use $q_i^+$ (positive pickup size) instead of the signed demand $\pm q_i$ because after $(p_r,d_r)$ contraction the signed values cancel to zero and the capacity constraint would degenerate; $q_i^+$ is a conservative upper bound since actual route load is sequence-dependent and never exceeds the summed pickup weight. Final CPDPTW feasibility is checked after per-cluster sequencing.

The conflict graph uses normalized demand pressure and time-window incompatibility,
\begin{equation}
w_{ij}=\alpha\frac{q_i^+q_j^+}{Q^2}+\beta\frac{\delta t_{ij}}{\Theta},
\label{eq:weights}
\end{equation}
where $\delta t_{ij}\geq0$ measures pairwise temporal incompatibility, for example separation of time-window midpoints or a failed-insertion penalty. The precise predictor can be learned or engineered; the quantum question is how its induced graph structure affects QRAO. In the full graph, most pairs receive nonzero weights, so $G=(V',E)$ is typically dense. We therefore introduce a sparsification level $s$ and form $G_s$ by either thresholding small $w_{ij}$ or retaining $k$ nearest neighbors in demand--time-window feature space. This step is not a cosmetic preprocessing choice; it controls whether QRAO compression is possible.

\section{Density Obstruction}
\label{sec:density}

QRAO cannot arbitrarily place adjacent variables on the same qubit. If two binary variables interact in the problem graph, assigning them to non-commuting Paulis on one qubit makes them impossible to determine simultaneously by a single rounding rule. Existing QRAO implementations therefore assign variables sharing a qubit only when they form an independent set in the problem graph. Achieved compression is governed by graph coloring and independent-set size.
\begin{equation}
\rho_{\mathrm{comp}}(G_s)=\frac{\text{number of binary variables}}{\text{number of relaxed qubits}}.
\label{eq:compression}
\end{equation}
For a complete graph, every independent set has size one and $\rho_{\mathrm{comp}}=1$; QRAO becomes a non-compressing relaxation. This is the density obstruction.

The obstruction is acute for CPDPTW. Equation~\eqref{eq:weights} creates application-derived dense graphs because most request pairs receive nonzero demand or time-window incompatibility. Even if the objective is sparsified, penalty-encoded constraints reintroduce cliques. A quadratic capacity penalty,
\begin{equation}
A\left(\sum_i q_i^+x_{i,c}-Q\right)^2,
\label{eq:penalty}
\end{equation}
contains all-pairs products $x_{i,c}x_{j,c}$ for each cluster $c$. Slack-variable encodings and unbalanced penalization can reduce qubit overhead or coefficient magnitude~\cite{montanez2024unbalanced}, but they do not remove the all-pairs graph density induced by the square. Thus constraint penalties directly attack the independent-set structure QRAO needs. One-hot compression stays at the nominal $3\times$ regardless of sparsification, so binary and higher-order binary optimization (HOBO) encodings give the conservative compression view; Section~\ref{sec:split} reports the collapse quantitatively.

This produces the central routing-specific trade-off.
\begin{equation}
\text{sparser }G_s
\Rightarrow
\begin{cases}
\text{larger independent sets},\\
\text{higher QRAO compression},\\
\text{smaller }\lambda\text{ and shallower circuits},\\
\text{larger clustering approximation error}.
\end{cases}
\label{eq:sparsity_tradeoff}
\end{equation}
The central question is not whether the graph should be sparse or dense, but where the operating point lies for CPDPTW.

\section{Sparse QRAO Repair}
\label{sec:repair}

The method separates three roles. The quantum component handles only the sparsified clustering objective, the classical component restores routing feasibility, and the noise model checks whether the sampled distribution remains useful on a calibrated backend. Keeping one-hot, capacity, and time-window constraints out of the Hamiltonian avoids clique-forming penalty terms, while repair ensures those constraints are not ignored.

\providecommand{\icreq}{\begin{tikzpicture}[baseline=-0.5ex,x=3mm,y=3mm,>=Latex]\fill (0,0) circle (0.5);\draw[->,shorten <=0.6, shorten >=0.6] (0,0) -- (3.2,0);\draw (3.2,0) circle (0.55);\end{tikzpicture}}
\providecommand{\icmerge}{\begin{tikzpicture}[baseline=-0.5ex,x=3mm,y=3mm,>=Latex]\fill (0,0.9) circle (0.45);\fill (0,-0.9) circle (0.45);\draw[->,shorten <=0.55, shorten >=0.55] (0,0.9) -- (2.8,0);\draw[->,shorten <=0.55, shorten >=0.55] (0,-0.9) -- (2.8,0);\fill (2.8,0) circle (0.6);\end{tikzpicture}}
\providecommand{\icdense}{\begin{tikzpicture}[baseline=-0.5ex,x=3mm,y=3mm]\foreach \a in {45,135,225,315} \fill (\a:1.7) circle (0.35);\draw (45:1.7) -- (135:1.7) -- (225:1.7) -- (315:1.7) -- cycle;\draw (45:1.7) -- (225:1.7);\draw (135:1.7) -- (315:1.7);\end{tikzpicture}}
\providecommand{\icsparse}{\begin{tikzpicture}[baseline=-0.5ex,x=3mm,y=3mm]\foreach \a in {45,135,225,315} \fill (\a:1.7) circle (0.35);\draw (45:1.7) -- (135:1.7);\draw (225:1.7) -- (315:1.7);\draw[gray!45,dashed,line width=0.25] (135:1.7) -- (225:1.7);\end{tikzpicture}}
\providecommand{\icqubit}{\begin{tikzpicture}[baseline=-0.5ex,x=3mm,y=3mm]\draw[line width=0.4] (0,0) circle (1.7);\node[font=\scriptsize] at (-0.75,0.5) {X};\node[font=\scriptsize] at (0.75,0.5) {Y};\node[font=\scriptsize] at (0,-0.8) {Z};\end{tikzpicture}}
\providecommand{\icmatrix}{\begin{tikzpicture}[baseline=-0.5ex,x=3mm,y=3mm]\draw[line width=0.4] (-1.6,-1.6) rectangle (1.6,1.6);\draw[gray!50] (-1.6,0) -- (1.6,0);\draw[gray!50] (0,-1.6) -- (0,1.6);\fill (0.8,0.8) circle (0.32);\fill (-0.8,-0.8) circle (0.32);\fill (-0.8,0.8) circle (0.25);\fill (0.8,-0.8) circle (0.25);\end{tikzpicture}}
\providecommand{\ichist}{\begin{tikzpicture}[baseline=-0.5ex,x=3mm,y=3mm]\fill (-1.8,-1.1) rectangle (-1.0,1.2);\fill (-0.4,-1.1) rectangle (0.4,0.4);\fill (1.0,-1.1) rectangle (1.8,-0.2);\draw[line width=0.35] (-2.1,-1.15) -- (2.1,-1.15);\end{tikzpicture}}
\providecommand{\icround}{\begin{tikzpicture}[baseline=-0.5ex,x=3mm,y=3mm]\draw[line width=0.4] (0,0) circle (1.6);\node[font=\scriptsize] at (0,0) {$\pm 1$};\end{tikzpicture}}
\providecommand{\icrepair}{\begin{tikzpicture}[baseline=-0.5ex,x=3mm,y=3mm,>=Latex]\draw[rounded corners=0.5,line width=0.35] (-2.1,-1.1) rectangle (0.4,1.1);\fill (-1.4,0.35) circle (0.32);\fill (-0.5,-0.35) circle (0.32);\fill (2.2,0.3) circle (0.32);\draw[->,shorten <=0.4, shorten >=0.4] (2.2,0.3) -- (0.0,0.3);\end{tikzpicture}}
\providecommand{\iccluster}{\begin{tikzpicture}[baseline=-0.5ex,x=3mm,y=3mm]\foreach \x/\c in {-2.2/red!65!black, 0/blue!60!black, 2.2/green!45!black}{\draw[rounded corners=0.35, draw=\c, line width=0.4] (\x-0.7,-0.85) rectangle (\x+0.7,0.85);\fill (\x,-0.3) circle (0.24);\fill (\x,0.3) circle (0.24);}\end{tikzpicture}}
\providecommand{\icroute}{\begin{tikzpicture}[baseline=-0.5ex,x=3mm,y=3mm,>=Latex]\foreach \x/\y in {-2.0/0.5, -0.7/-0.55, 0.7/0.5, 2.0/-0.5} \fill (\x,\y) circle (0.3);\draw[line width=0.35,->,shorten <=0.35, shorten >=0.35] (-2.0,0.5) -- (-0.7,-0.55) -- (0.7,0.5) -- (2.0,-0.5);\end{tikzpicture}}
\providecommand{\icgap}{\begin{tikzpicture}[baseline=-0.5ex,x=3mm,y=3mm]\draw[line width=0.35] (-2.0,-1.15) -- (2.0,-1.15);\fill (-1.2,-1.15) rectangle (-0.4,1.05);\fill[gray!55] (0.4,-1.15) rectangle (1.2,0.15);\end{tikzpicture}}
\providecommand{\icwave}{\begin{tikzpicture}[baseline=-0.5ex,x=3mm,y=3mm]\draw[line width=0.5] (-2.1,0) sin (-1.05,0.8) cos (0,0) sin (1.05,-0.8) cos (2.1,0);\end{tikzpicture}}
\providecommand{\icmetric}{\begin{tikzpicture}[baseline=-0.5ex,x=3mm,y=3mm,>=Latex]\draw[line width=0.4] (-1.5,0) arc (180:0:1.5);\draw[line width=0.5,->,shorten >=0.2] (0,0) -- (55:1.35);\fill (0,0) circle (0.25);\end{tikzpicture}}
\providecommand{\icguard}{\begin{tikzpicture}[baseline=-0.5ex,x=3mm,y=3mm,>=Latex]\fill[gray!25,rounded corners=0.3] (-1.3,-0.9) rectangle (1.3,0.9);\draw[->,line width=0.45, shorten <=0.2, shorten >=0.2] (-1.7,-0.6) .. controls (-1.0,1.6) and (1.0,1.6) .. (1.7,-0.6);\end{tikzpicture}}
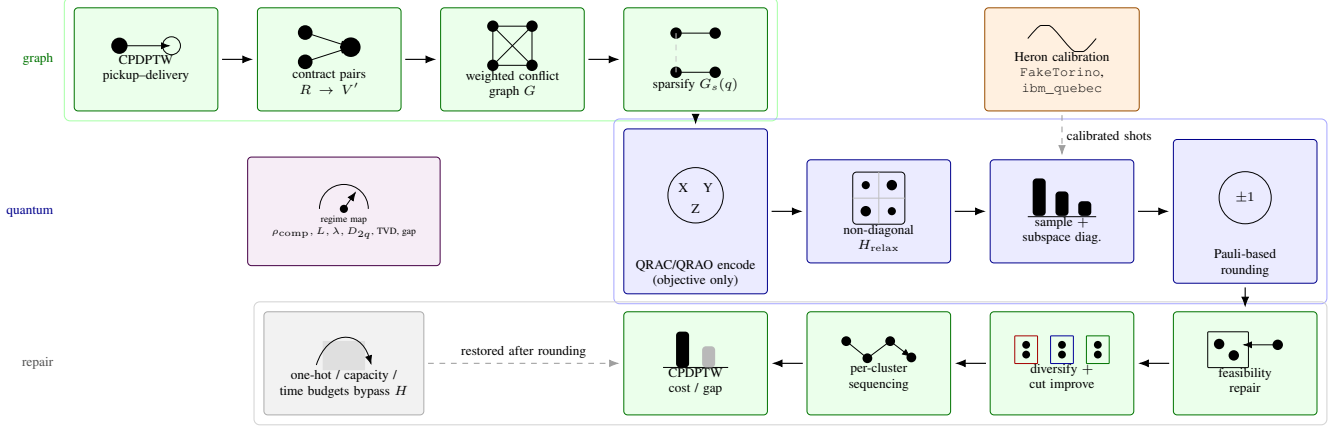
\begin{figure*}[t]
\centering
\scriptsize
\resizebox{0.99\textwidth}{!}{%
\begin{tikzpicture}[
  >=Latex,
  x=27mm, y=28mm,
  stage/.style={draw, rounded corners=2pt, align=center, minimum height=19mm, inner xsep=3.5pt, inner ysep=3.5pt, text width=24mm, font=\scriptsize},
  qstage/.style={stage, fill=blue!8, draw=blue!55!black},
  cstage/.style={stage, fill=green!8, draw=green!45!black},
  nstage/.style={stage, fill=orange!12, draw=orange!60!black, text width=26mm},
  guard/.style={stage, fill=gray!10, draw=gray!65, text width=27mm},
  metric/.style={stage, fill=violet!7, draw=violet!55!black, text width=33mm, minimum height=20mm, font=\tiny},
  lane/.style={draw, rounded corners=3pt, inner xsep=5pt, inner ysep=5pt, line width=0.35pt},
  flow/.style={->, line width=0.6pt, shorten >=2pt, shorten <=2pt},
  side/.style={->, dashed, line width=0.55pt, draw=gray!75, shorten >=2pt, shorten <=2pt},
  lab/.style={align=center, font=\scriptsize}
]
\node[cstage] (pd)       at (0, 1.0)    {\icreq\\CPDPTW\\pickup--delivery};
\node[cstage] (contract) at (1.25, 1.0) {\icmerge\\contract pairs\\$R\!\to\!V'$};
\node[cstage] (graph)    at (2.50, 1.0) {\icdense\\weighted conflict\\graph $G$};
\node[cstage] (sparse)   at (3.75, 1.0) {\icsparse\\sparsify $G_s(q)$};

\node[qstage] (qrao)  at (3.75, 0.0) {\icqubit\\QRAC/QRAO encode\\(objective only)};
\node[qstage] (ham)   at (5.00, 0.0) {\icmatrix\\non-diagonal\\$H_{\mathrm{relax}}$};
\node[qstage] (sqd)   at (6.25, 0.0) {\ichist\\sample $+$\\subspace diag.};
\node[qstage] (round) at (7.50, 0.0) {\icround\\Pauli-based\\rounding};

\node[cstage] (feas)  at (7.50,-1.0) {\icrepair\\feasibility\\repair};
\node[cstage] (imp)   at (6.25,-1.0) {\iccluster\\diversify $+$\\cut improve};
\node[cstage] (route) at (5.00,-1.0) {\icroute\\per-cluster\\sequencing};
\node[cstage] (score) at (3.75,-1.0) {\icgap\\CPDPTW\\cost / gap};

\node[guard]  (constraints) at (1.35,-1.0) {\icguard\\one-hot / capacity /\\time budgets bypass $H$};
\node[nstage] (noise)       at (6.25, 1.0) {\icwave\\Heron calibration\\\texttt{FakeTorino}, \texttt{ibm\_quebec}};
\node[metric] (metrics)     at (1.35, 0.0) {\icmetric\\regime map\\$\rho_{\rm comp},L,\lambda,D_{2q}$, TVD, gap};

\node[fit=(pd)(contract)(graph)(sparse), lane, draw=green!35] (lanegraph) {};
\node[fit=(qrao)(ham)(sqd)(round), lane, draw=blue!35] (lanequantum) {};
\node[fit=(constraints)(score)(route)(imp)(feas), lane, draw=gray!35] (lanerepair) {};
\node[font=\scriptsize, text=green!45!black, anchor=east] at (-0.6,1.0) {graph};
\node[font=\scriptsize, text=blue!55!black, anchor=east] at (-0.6,0.0) {quantum};
\node[font=\scriptsize, text=gray!65!black, anchor=east] at (-0.6,-1.0) {repair};

\draw[flow] (pd) -- (contract);
\draw[flow] (contract) -- (graph);
\draw[flow] (graph) -- (sparse);
\draw[flow] (sparse.south) -- (qrao.north);
\draw[flow] (qrao) -- (ham);
\draw[flow] (ham) -- (sqd);
\draw[flow] (sqd) -- (round);
\draw[flow] (round.south) -- (feas.north);
\draw[flow] (feas) -- (imp);
\draw[flow] (imp) -- (route);
\draw[flow] (route) -- (score);

\draw[side] (constraints.east) -- (score.west)
    node[lab, midway, above] {restored after rounding};
\draw[side] (noise.south) -- (sqd.north)
    node[lab, midway, right] {calibrated shots};
\end{tikzpicture}
}
\caption{Sparse QRAO pipeline with two-phase repair.} 
\label{fig:pipeline}
\end{figure*}

The relaxed Hamiltonian is built only from the sparsified objective in~\eqref{eq:maxkcut}. The experiments use the fixed shallow SKQD-style preparation defined in Section~\ref{sec:characterization}; transferred low-parameter ansatzes as in SQOA-QR~\cite{sqoa2025} are compatible with the same pipeline. Because $H_{\mathrm{relax}}$ contains non-commuting Pauli terms, projected subspace diagonalization is non-trivial and is no longer equivalent to best-of-shots. The decoded assignments are obtained by Pauli-expectation rounding or the alternative single-qubit rounding rule of~\cite{fuller2024qrao}. Neither rounding rule guarantees one-hot assignment or per-cluster capacity and time-budget proxy feasibility, so repair is required.

The repair operator runs in two phases. The feasibility phase enforces one cluster per request, capacity, and a time-budget proxy by moving requests from overloaded clusters to admissible clusters with minimal objective damage. If needed, a pressure-ordered pack-from-scratch fallback restores feasibility. The improvement phase seeds empty clusters with high-degree admissible requests and then applies cut-improving single-node reassignments while preserving all budgets. This avoids the feasible but useless all-in-one-cluster solution. On the Heron hardware slice in Table~\ref{tab:calibration}, repair reduces the mean rounded gap from $0.493{\pm}0.035$ to $0.0016{\pm}0.0012$ using $4.47{\pm}0.42$ moves.

After repair, each cluster is sequenced by exact dynamic programming when small enough, or by OR-Tools/ALNS otherwise. The final score is the true CPDPTW route cost, not the relaxed Max-$K$-Cut energy.

The end-to-end pipeline takes CPDPTW requests, fleet size $K$, sparsification grid $\mathcal{S}$, and shot budget $B$. It first
(1) contracts each pickup--delivery pair into a supernode set $V'$ and builds the weighted conflict graph $G$ using~\eqref{eq:weights}.
For each $s\in\mathcal{S}$, it then
(2) constructs $G_s$ by thresholding or $k$-NN pruning;
(3) expands Max-$K$-Cut variables in one-hot or binary form and QRAO-colors the variable graph, recording $\rho_{\mathrm{comp}}(G_s)$;
(4) builds non-diagonal $H_{\mathrm{relax}}(G_s)$ from the objective only;
(5) prepares the fixed shallow sampling state used in the experiments and draws $B$ samples;
(6) diagonalizes $H_{\mathrm{relax}}$ in the sampled subspace and rounds to a cluster assignment by Pauli-expectation measurement;
(7) runs the two-phase repair;
(8) sequences each repaired cluster and evaluates the true CPDPTW cost.
The output is a regime map over sparsification level, compression, circuit resources, repair cost, and routing gap.

\section{Characterization Protocol}
\label{sec:characterization}

The characterization is a resource and robustness study, not an advantage claim. For each sparsification level we log
\begin{equation}
s \mapsto \left(|E_s|, \rho_{\mathrm{comp}}, n_q, L, \lambda, D_{2q}, \eta_{\mathrm{sample}}, g_{\mathrm{route}}\right),
\label{eq:chain}
\end{equation}
where $|E_s|$ is graph size, $n_q$ relaxed qubits, $L$ Pauli-term count, $\lambda=\sum_j|\alpha_j|$, $D_{2q}$ transpiled two-qubit depth on IBM Heron heavy-hex, $\eta_{\mathrm{sample}}$ a sampling-efficiency statistic, and $g_{\mathrm{route}}$ the final routing gap.

For backend $b$ and calibration timestamp $\tau$, we build a calibrated noise model $\mathcal{N}_{b,\tau}$ from gate errors, $T_1/T_2$ data, and readout errors, using depolarizing, thermal-relaxation, and measurement channels. Selected runs are evaluated under ideal Aer, calibrated Aer, and hardware execution, with TVD used to compare the sampled distributions. The ALNS-derived noise score is only an ordering prior for shot allocation; without held-out cross-backend validation it is not a calibrated predictor.

All sampling experiments use the same fixed SKQD-style preparation. The initial state is $\ket{\psi_0}=\ket{0}^{\otimes n_q}$, and the Krylov states are
\begin{equation}
\ket{\psi_k}=e^{-iH_{\rm relax} k\Delta t}\ket{\psi_0},\quad k=0,\ldots,r-1,
\end{equation}
implemented with a one-step Lie--Trotter \texttt{PauliEvolutionGate}. The step size is $\Delta t=\pi/\|H_{\rm relax}\|$ when the norm is available, with $\pi/\lambda$ as the fallback. Thus $r$ is the number of sampled time points, not an ansatz depth parameter; at $r=1$ the only prepared state is $\ket{0}^{\otimes n_q}$, so the reported energy comes only from the Hamming-1 neighbor expansion. Any gain for $r>1$ comes from the time-evolved samples. The sampled subspace at each $r$ is the union of measured bitstrings from all $r$ circuits, expanded by Hamming-1 neighbors of the top 32 sampled configurations before diagonalizing $H_{\rm relax}$.

Baselines are classical clustering with OR-Tools/ALNS sequencing and diagonal best-of-shots on the same sparsified objective. A randomized-rounding surrogate is deferred, so success here means a calibrated regime map rather than unconditional superiority. QSVT and LCU are analytic resource lenses: for a filter $P_d(H/\lambda)$, $d\sim \lambda/\Delta$. Sparsification and repair-after-rounding reduce $\lambda$ by removing objective edges and clique-forming penalties; Fej\'er-filter constructions give related finite-depth, finite-shot guarantees when constraints are handled outside penalties~\cite{onah2025fejer}.

\section{Experiments}
\label{sec:split}

\begin{figure}[t]
\centering
\includegraphics[width=\columnwidth]{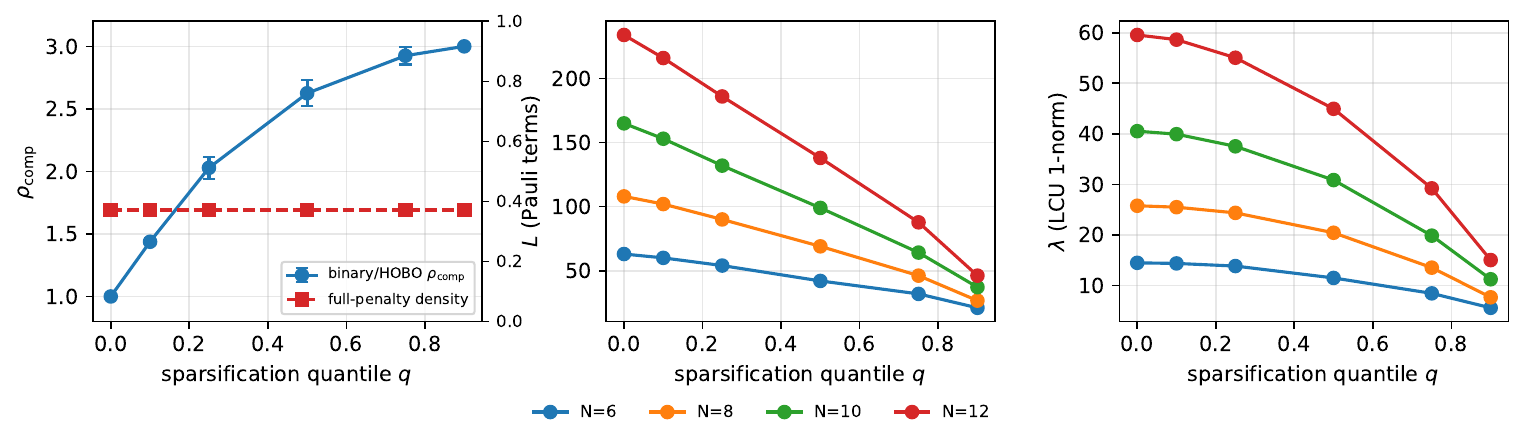}
\caption{Encoding resources versus sparsification level. Left: binary/HOBO compression recovers as the objective graph is sparsified, while full penalty encoding keeps interaction density pinned. Middle/right: Pauli-term count $L$ and LCU norm $\lambda$ both drop with the quantile $q$. Statistics use $N\in\{6,8,10,12\}$ and eight seeds.}
\label{fig:resource_curves}
\end{figure}

We report focused experiments that test the structural claims. The first experiment is a classical graph analysis. From each CPDPTW instance we construct the full conflict graph using~\eqref{eq:weights} and generate a sweep of sparsified graphs $G_s$ using thresholding and $k$-nearest-neighbor rules. For each $G_s$ we compute graph density, independent-set statistics, achieved QRAO compression, and the LCU 1-norm of the relaxed objective. On instances of size $N\in\{6,8,10,12\}$, the LCU norm $\lambda$ drops with the sparsification quantile (from $\lambda\approx 26$ at $q=0$ to $\lambda\approx 8$ at $q=0.9$ for $N=8$) while one-hot QRAO compression stays at the nominal $3\times$; binary/HOBO compression, in contrast, collapses from $3\times$ at $q=0.9$ to $1\times$ at $q=0$. Figure~\ref{fig:resource_curves} plots the three resource axes.

The second experiment quantifies the constraint-density claim. We compare three Hamiltonian choices at $N\!=\!12$: objective-only sparsified Max-$K$-Cut with repair after rounding, sparsified objective plus one-hot penalties, and sparsified objective plus one-hot and capacity/time penalties. Objective-only interaction density falls with $q$ as expected ($0.31\!\to\!0.04$ across $q\!\in\![0,0.9]$), and binary/HOBO compression tracks it ($1.00\!\times\!\to 3.00\!\times$). Adding penalties pins the interaction density. One-hot penalties raise it by $+0.06$ to $+0.09$ at every $q$, and full penalties lock it at $0.371$ independent of $q$. These numbers are the empirical version of eq.~\eqref{eq:penalty}. Unbalanced penalization 
reduces coefficient scale and hence $\lambda$, but it does not remove the clique structure created by quadratic constraints.

\begin{figure*}[t]
\centering
\includegraphics[width=0.8\textwidth]{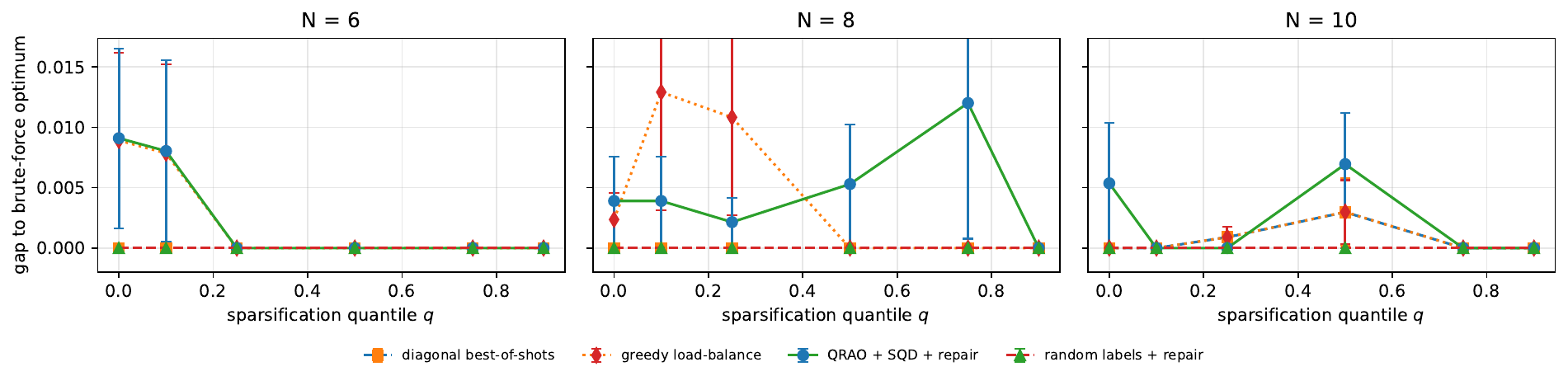}
\caption{Method comparison under full two-phase repair, $N\!\in\!\{6,8,10\}$, mean $\pm$ SE over eight seeds with brute-force oracle gaps. Random-labels and diagonal best-of-shots overlap at gap $\approx 0$. QRAO+SQD+repair can sit slightly above them because Pauli rounding often starts from a collapsed assignment and deterministic local repair can settle in a worse basin than repair from random labels. This near-zero collapse is expected, so the non-diagonal contribution is measured in the feasibility-only decoder ablation. $N=12$ is omitted from the panel for space; the decoder ablation in the text covers it.}
\label{fig:method_comparison}
\end{figure*}

The third experiment compares QRAO+SQD+repair with diagonal best-of-shots, random-labels+repair, and greedy clustering under a matched shot budget (Fig.~\ref{fig:method_comparison}). Under full two-phase repair all four converge to gap $\leq 0.02$, with diagonal best-of-shots tracking random-labels as predicted by eq.~\eqref{eq:best_shots}. Non-diagonality is measured first at the energy level: at $N\!=\!8$, SQD ground energy drops from $-3.20{\pm}0.55$ at $r\!=\!1$ to $-5.51{\pm}1.19$ at $r\!=\!5$ ($10$ seeds). Assignment quality is decoder-dependent. With a fixed top-32 eigenvector rerank, feasibility-only gains over random labels are $48\%$, $37\%$, $9\%$, and $3\%$ for $N=6,8,10,12$. Widening the rerank to 512 candidates changes the $N=10,12$ gaps from $0.281$ and $0.265$ to $0.131{\pm}0.024$ and $0.173{\pm}0.018$, so the apparent fade is partly a decoder-budget effect. Best feasible candidates appear deep in the $|\alpha_x|^2$ ordering, with median ranks $340$ at $N=10$ and $785$ at $N=12$; this explains why the widened budget helps more at $N=10$ than at $N=12$.

\begin{figure}[t]
\centering
\includegraphics[width=\columnwidth]{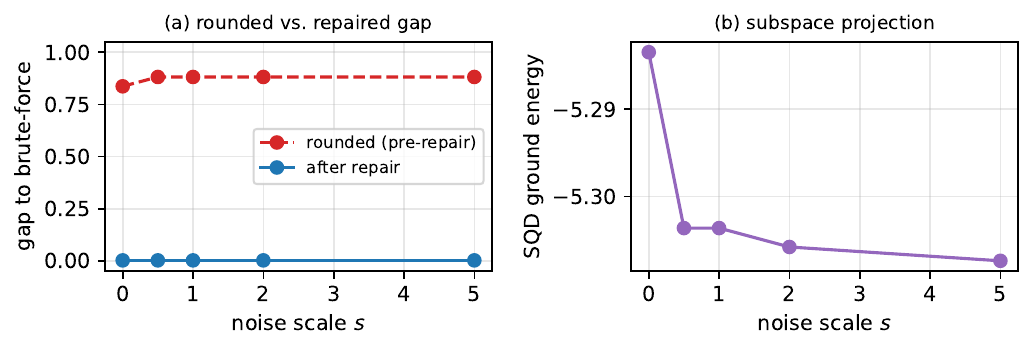}
\caption{Noise disambiguation at $N\!=\!8$, $K\!=\!3$, 8 seeds. (a) Rounded gap stays near $0.88$ because Pauli rounding collapses to one cluster, while repaired gap is near zero. (b) SQD ground energy varies by less than $0.03$ ($-5.28\!\to\!-5.31$) across $s\!\in\!\{0,\ldots,5\}$; the small visual changes are within seed-level variation.}
\label{fig:noise_sensitivity}
\end{figure}

The fourth experiment is a noise-robustness study. Fixing the best sparsification level from the third experiment, we sweep a synthetic noise scale $s\in\{0,1,2,5\}$ over an IBM Heron-derived noise model (\texttt{FakeTorino}) at $N=8$. Figure~\ref{fig:noise_sensitivity}(b) shows that the SQD ground-state energy varies by less than $0.03$ across the full sweep ($-5.283\to-5.307$ over $8$ seeds). The subspace projection is therefore stable under this calibrated noise model. Re-running the subspace-argmax decoder at $s\!=\!5$ over $10$ seeds gives feasibility-only gap $0.231$, essentially identical to the noiseless $0.23$, so decoding from the SQD eigenvector also survives the noise sweep. Panel (a) shows the limitation. Rounded gap stays near $0.88$ because the fixed shallow Krylov ansatz remains far from the relaxed ground state, so Pauli-expectation rounding collapses the assignment before repair can act. The classical top-1 shot share drops from $0.46$ to $0.19$ and Gini from $0.85$ to $0.65$ under noise, but these raw shot statistics do not determine the post-repair gap.

Finally, we ran the full sparse-QRAO pipeline on IBM Heron \texttt{ibm\_quebec}. The study contains a validated $108$-row grid combining ideal Aer, calibrated \texttt{FakeTorino}, and real hardware records. It spans $N\in\{4,6,8,10,12\}$, $K\in\{2,3,4\}$, sparsification $q\in\{0,0.25,0.5,0.75\}$, Krylov dimensions $r\in\{1,3,5,7\}$, and both $(3,1)$ and $(2,1)$ QRAC encodings. Every hardware row produces a feasible repaired assignment; the mean hardware gap is $0.0016$ and the worst observed gap is $0.0382$. Distribution drift grows with size (Table~\ref{tab:calibration}), but repair stabilizes the decoded clustering quality.

\begin{table}[t]
\centering
\caption{Final Heron hardware summary. Observed TVD compares real \texttt{ibm\_quebec} sampling with ideal QRAO sampling after identical decoding. The null floor estimates finite-shot TVD between two ideal samples; observed TVD is $1.8$--$4.3\times$ larger.}
\label{tab:calibration}
\footnotesize
\begin{tabular}{lccccc}
\toprule
Slice & hw & obs.\ TVD & null floor & obs./floor & mean gap \\
\midrule
all hardware & 36 & 0.279 & 0.114 & 2.44 & 0.0016 \\
$N\!=\!4$ & 3 & 0.174 & 0.041 & 4.28 & 0.0000 \\
$N\!=\!6$ & 14 & 0.185 & 0.078 & 2.38 & 0.0000 \\
$N\!=\!8$ & 12 & 0.324 & 0.133 & 2.43 & 0.0032 \\
$N\!=\!10$ & 3 & 0.391 & 0.195 & 2.01 & 0.0000 \\
$N\!=\!12$ & 4 & 0.463 & 0.254 & 1.82 & 0.0049 \\
\bottomrule
\end{tabular}
\end{table}

At $N\!=\!12$, sparsification reduces $L$ from $186$ to $87$ and TVD from $0.568$ to $0.279$ as $q$ moves $0.25\!\to\!0.75$. The ALNS prior of~\cite{moosavifarooq2025} predicts mean TVD $0.614$ versus observed $0.279$, so it is not a calibrated QRAO noise model; its ranking survives with Spearman $0.854$ and Pearson $0.801$. Shot coverage $3000/K^N$ is $46\%$, $5\%$, and $0.6\%$ at $(N,K)\!=\!(8,3),(10,3),(12,3)$; decoded rounded-label agreement with ideal sampling is $67\%$ overall and repaired-label agreement is $78\%$ ($75\%$ at $N=12$). Across all hardware rows, repair reduces the mean rounded gap from $0.493{\pm}0.035$ to $0.0016{\pm}0.0012$ with $4.47{\pm}0.42$ moves. Near-zero classical-baseline gap at $N\!\leq\!8$ is expected, while the near-zero hardware gap at $N\!\in\!\{10,12\}$ is the main evidence for robust repair.

\section{Conclusion and Future Work}
\label{sec:conclusion}

Diagonal SQD is redundant with best-of-shots for routing. Quantum relaxation supplies non-diagonality, but CPDPTW graph density obstructs compression. Two-phase repair restores feasibility and routing quality. Under strong repair, the classical pass alone can reach the optimum, so the quantum contribution is structural and regime-dependent. The supporting evidence is that SQD energy improves with Krylov dimension, drifts by a small amount under FakeTorino noise, sustains about threefold architectural qubit compression, and gives a decoder-dependent signal that improves when the rerank budget grows.

This work provides three reusable baselines: an achieved-compression curve versus sparsification on CPDPTW graphs, a calibrated Heron record with a TVD null floor, and decoder-budget scaling for subspace-argmax gains.

Future work includes alternative single-qubit rounding rules, learned graph pruning, QSVT and Fej\'er filtering, larger-instance hardware, and matched-budget QAOA and CVaR baselines. 


\end{document}